\documentclass[pra,twocolumn,a4paper,groupedaddress,showpacs,floatfix]{revtex4-1}
\usepackage{graphicx,amsmath,amssymb,amsfonts,dsfont}
\newcommand{\mf}[1]{\boldsymbol{#1}}
\newcommand{\ket}[1]{\ensuremath{|#1\rangle}}
\newcommand{\mc}[1]{\ensuremath{\mathcal{#1}}}
\newcommand{\bra}[1]{\ensuremath{\langle #1 |}}
\newcommand{\braket}[2]{\ensuremath{\langle #1 | #2 \rangle}}
\newcommand{\ve}{\varepsilon}

\newcommand{\mean}[1]{\ensuremath{ \langle #1  \rangle}}

\usepackage{color}

\begin{document}

\title{Abelian and non-Abelian gauge fields in dipole-dipole interacting Rydberg atoms}

\author{Martin Kiffner${}^{1,2}$}
\author{Wenhui Li${}^{1,3}$}
\author{Dieter Jaksch${}^{2,1}$}

\affiliation{Centre for Quantum Technologies, National University of Singapore, 3 Science Drive 2, Singapore 117543${}^1$}
\affiliation{Clarendon Laboratory, University of Oxford, Parks Road, Oxford OX1 3PU, United Kingdom${}^2$}
\affiliation{Department of Physics, National University of Singapore, 117542, Singapore${}^3$}

\pacs{03.65.Vf,03.75.-b,32.80.Rm}








\begin{abstract}
We show that  the dipole-dipole interaction between two Rydberg atoms can lead to 
substantial Abelian and non-Abelian gauge fields acting on the relative motion of the two atoms. 
We demonstrate how the gauge fields can be evaluated by numerical techniques.
In the case of adiabatic motion in a single internal state, we show that the gauge fields 
give rise to a magnetic field that results in a Zeeman splitting of 
the rotational states. In particular, the ground state of a molecular potential well is given by the 
first excited rotational state. 
We find that our system realises a synthetic spin-orbit coupling  where the relative 
atomic motion couples to  two internal two-atom states. 
The associated gauge fields are  non-Abelian. 
\end{abstract}

\maketitle

%
%
\section{Introduction}
Gauge theories are of fundamental importance in modern theoretical physics. Examples  
are given by classical and quantum electrodynamics and the standard model of elementary particle physics. 
In addition, it has been realised that Abelian and non-Abelian gauge fields arise in the adiabatic 
evolution of quantum mechanical systems~\cite{wilczek:84}. 
This concept was   applied to molecules in~\cite{moody:86}, where the realisation of magnetic monopoles 
is discussed.
However, the simulation of gauge fields with conventional molecules is impractical for several reasons. 
First,  the  gauge fields arise from those  terms that are usually neglected in the ubiquitous 
Born-Oppenheimer (BO) approximation~\cite{born:27}.  
Since the BO approximation is very well satisfied  in many systems due to 
the huge mass difference between electrons and nuclei, the desired gauge field effects are usually 
very small. Second, the experimental observation is considerably hampered by the small 
size of conventional molecules. Third, not all gauge fields  give rise to magnetic fields, and 
it is very challenging to engineer non-trivial gauge fields via external manipulations of 
the electronic levels. 
For these reasons, other systems for the simulation of artificial gauge fields have been explored. 
In particular, tremendous experimental 
and theoretical effort has been made to simulate gauge fields
with cold  atoms~\cite{ruseckas:05,dalibard:11,dum:96,lin:09,lin:09n,lin:11,aidelsburger:11,jaksch:03,struck:11,struck:12,jimenez_garcia:12,hauke:12} 
in  tailored electromagnetic fields. 
The exaggerated properties of Rydberg atoms~\cite{gallagher:ryd}  give rise to various 
exceptional phenomena in quantum optics~\cite{pritchard:12} and quantum information science~\cite{saffman:10}. 
For example, it is possible to form  exotic molecules that are orders of magnitude larger than 
naturally occurring molecules. 
So-called trilobite molecules are comprised of one Rydberg atom and one ground state atom, where 
the bond is mediated by the interaction of the Rydberg electron with the ground state atom~\cite{greene:00,bendkowsky:09}. 
On the other hand, Rydberg macrodimers consist of  two Rydberg atoms with non-overlapping electron 
clouds~\cite{boisseau:02,samboy:11,kiffner:12,overstreet:09}, and 
the binding potential arises from the dipole-dipole interaction. 
Since typical internuclear spacings  exceed $1\,\mu\text{m}$, the experimental observation of 
Rydberg-Rydberg correlations becomes feasible~\cite{schwarzkopf:11}. 
Moreover, several features like 
the internuclear spacing and  the angular dependence of the binding potential can be controlled in some 
Rydberg macrodimer schemes.  
%
\begin{figure}[t!]
\includegraphics[width=6cm]{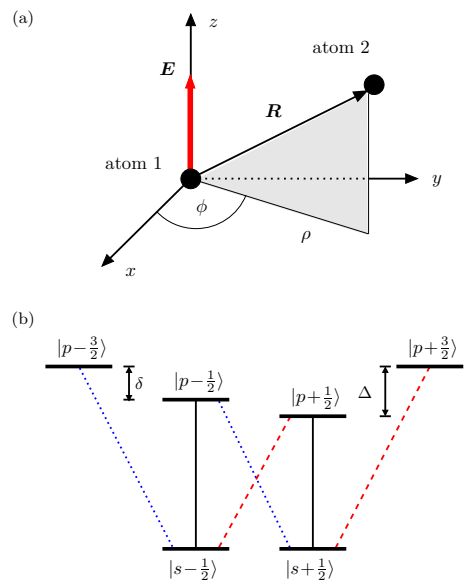}
\caption{\label{fig1}
(Color online) (a) The system under consideration
consists of two Rydberg atoms. $\mf{R}$ is the relative position  of atom 2 with respect to atom 1. 
An external electric field $\mf{E}$ is applied
in the  $z$ direction. $\rho$ is the distance of atom 2 from the $z$ axis.
(b) Internal level structure of each Rydberg atom.   The Stark shifts $\delta\equiv W_{p -1/2}-W_{p -3/2}$ 
and $\Delta\equiv W_{p +1/2}-W_{p+ 3/2}$ are negative. We assume $\delta\not=\Delta$. 
The dipole transitions  indicated by solid, blue dotted 
and red dashed lines couple to $\pi$, $\sigma^-$ and $\sigma^+$ polarised fields, respectively. }
\end{figure}
%

%
Here we show that the dipole-dipole interaction between two Rydberg atoms gives rise to sizeable Abelian and 
non-Abelian gauge fields acting on the relative motion of the two atoms. 
Our system is several orders of magnitude larger than conventional molecules, and non-trivial gauge fields 
can be engineered via the exquisite ability to manipulate and control individual Rydberg atoms. 
More specifically, our system shown in Fig.~\ref{fig1} 
is based on the Rydberg macrodimer proposal in~\cite{kiffner:12}, but here we consider an 
asymmetric Stark shift which breaks the  symmetry of the system.  
Furthermore, it involves only a relatively small number of atomic states which 
facilitates the calculation of artificial gauge fields. 
We show that the dipole-dipole interaction gives rise to an artificial magnetic field for adiabatic motion 
in a donut-shaped potential well. 
This magnetic field results in a Zeeman shift of rotational states. 
In particular, the molecular ground state is given by the first excited rotational state. 
Furthermore, we show that the gauge fields can induce  non-adiabatic transitions near an avoided crossing where 
two internal  states become near-degenerate. Our system thus realises a synthetic spin-orbit coupling, and 
the associated gauge fields are strongly  non-Abelian.  
This paper is organised as follows. We give a detailed description of our system and  
the theoretical framework of artificial gauge fields in Sec.~\ref{model}. Some details are 
deferred to Appendix~\ref{internal}. 
In Sec.~\ref{abelian} we assume that the relative atomic motion is confined to a donut-shaped potential 
well and evaluate the corresponding Abelian vector potential. We find that it gives rise to a position-dependent 
magnetic field and calculate the first excited bound states of the potential well.  
In Sec.~\ref{nonab} we show that the relative atomic motion can induce non-adiabatic transitions between  two internal states 
exhibiting an avoided crossing. 
With the techniques described in Appendices~\ref{fields} and~\ref{curvature}  we evaluate 
the corresponding gauge fields and show that they are non-Abelian. 
Finally, we  conclude with  a summary and outlook in Sec.~\ref{summary}. 
%
\section{Model \label{model}} 
The geometry of the two atom system under consideration is shown in Fig.~\ref{fig1}a. 
The vector $\mf{R}$ is the relative position  of atom 2 with respect to atom 1. We are interested in the 
relative motion of the two Rydberg atoms that are coupled via the dipole-dipole interaction.
The total Hamiltonian for the relative motion of the two atoms and the internal states is 
\begin{align}
H = H_{\text{R}} + H_{\text{int}},
\end{align}
where 
\begin{align}
H_{\text{R}} = \frac{\mf{p}^2}{2\mu}
\end{align}
is the kinetic energy of the relative motion and $\mu$ is the reduced mass of the two Rydberg atoms. We omit the 
centre-of-mass motion which is uniform. 
The Hamiltonian describing the internal degrees of freedom is given by 
\begin{align}
H_{\text{int}} = H_{\text{S}} + V_{\text{dd}},
\label{Ht}
\end{align}
where $H_{\text{S}}$ describes the internal levels of the two uncoupled atoms and $V_{\text{dd}}$ is the dipole-dipole interaction 
 (for details see Appendix~\ref{internal}).
In each Rydberg atom we consider two angular momentum multiplets. The lower $ns_{1/2}$ states have total angular momentum $J=1/2$, 
and the excited multiplet is comprised of  $np_{3/2}$ states with total angular momentum $J=3/2$. We specify the individual 
atomic states by their orbital angular momentum $\ell$ and azimuthal total angular momentum $m_j$, 
i. e. as $|\ell m_j\rangle$. 
A DC electric field $\mf{E}$ in the $z$ direction  defines the quantisation axis and gives rise to 
Stark shifts of the magnetic sublevels. We assume that the Stark shifts are different  in the $m_j>0$ 
and $m_j<0$ manifolds, which could be achieved by inducing additional AC stark shifts, see Appendix~\ref{internal}. 
For simplicity  we focus on the level scheme shown in Fig.~\ref{fig1}(b), where the asymmetry is characterised 
by a single parameter $\Delta/\delta$ with $\delta\equiv W_{p -1/2}-W_{p -3/2}$ and $\Delta\equiv W_{p +1/2}-W_{p+ 3/2}$. 
The total  internal state space of the two atoms  is spanned by 36 states. Here we focus on 
the subspace spanned by the $nsnp$ states where one atom is in a $ns_{1/2}$ state and the other 
in a $np_{3/2}$ state. This subspace has dimension $N=16$ and is spanned by two sets of 8 basis states given by 
$\ket{s m_{1/2},p m_{3/2}}$ and  $\ket{p m_{3/2},s m_{1/2}}$. It is convenient to classify these two-atom states with respect to 
the total azimuthal angular momentum $M$. There are eight states with $|M|=1$, four with $M=0$ and four with $|M|=2$. 
The $nsnp$ states are eigenstates of the Hamiltonian $H_{\text{S}}$, but not of 
the dipole-dipole interaction $V_{\text{dd}}$. Since the dipole-dipole interaction depends on the magnitude 
and orientation of the separation vector $\mf{R}$, the eigenstates and eigenvalues of $H_{\text{int}}$ will depend on 
$\mf{R}$, too. In order to account for the azimuthal symmetry of the system, we express $\mf{R}$ in terms 
of cylindrical coordinates, 
\begin{align}
\mf{R} = (\rho \cos\phi,\rho \sin\phi, z). 
\end{align}
For every value of $\mf{R}$ we introduce a set of orthonormal 
eigenstates of the Hamiltonian $H_{\text{int}}$, 
\begin{align}
H_{\text{int}} \ket{\psi_i(\mf{R})} = \epsilon_i (\mf{R}) \ket{\psi_i(\mf{R})}, \label{eigen}
\end{align}
where $\epsilon_i (\mf{R})$ are the corresponding eigenvalues. Due to the azimuthal symmetry of the problem, 
the eigenvalues depend only on $z$ and $\rho$, but not on $\phi$~\cite{kiffner:07}. 
The full quantum state of the two-atom system is thus described by the state vector 
\begin{align}
\ket{\Psi} = \sum\limits_{i=1}^{N}\int \text{d}^3R \; \alpha_i(\mf{R}) \ket{\psi_i(\mf{R})}\otimes\ket{\mf{R}} , 
\label{genstate}
\end{align}
where the sum runs over the $N=16$ eigenstates of the $nsnp$ subspace. 
We follow the procedure described in~\cite{dum:96,ruseckas:05,dalibard:11} and derive from Eqs.~(\ref{Ht}) 
and~(\ref{genstate}) an effective Schr\"odinger equation for the wavefunctions $\mf{\alpha}=(\alpha_1,\ldots,\alpha_N)$.   
We obtain 
\begin{align}
\imath\hbar \partial_t \mf{\alpha} = \left[\frac{1}{2\mu}(\mf{p}\mathds{1} -\mf{A})^2 + V \right] \mf{\alpha}, 
\label{s1}
\end{align}
where $V$ and $\mf{A}$ are $N\times N$ matrices that  are defined as 
\begin{align}
V_{kl} & = \delta_{kl} \epsilon_k (\mf{R}) , \label{vdef} \\
\mf{A}_{kl} &  = \imath \hbar \bra{\psi_k(\mf{R})}\nabla\ket{\psi_l(\mf{R})} , \label{adef}
\end{align}
and $\delta_{kl}$ is the Kronecker delta. 
Note that each matrix element $\mf{A}_{kl}$ is a 3-column vector. 
In the following, $A^{(s)}$ denotes the matrix comprised of matrix elements $A_{nm}^{(s)}=\mf{A}_{nm}\cdot \mf{e}_s$, 
where $\mf{e}_s$ is a 3-column unit vector and the dot denotes the (real) scalar product. We use $s\in\{1,2,3\}$ for the Cartesian unit vectors 
as well as $s\in\{\rho,z,\phi\}$ with   
\begin{align}
& \mf{e}_{\rho}=\cos\phi\mf{e}_1 + \sin\phi\mf{e}_2, \quad \mf{e}_{z}=\mf{e}_{3}, \\
& \mf{e}_{\phi}=-\sin\phi\mf{e}_1 + \cos\phi\mf{e}_2\,.
\end{align}
The same notation convention applies to other objects of the same structure as $\mf{A}$. 

The off-diagonal terms $|\mf{A}_{kl}|$ ($k\not=l$) can induce transitions between two potential 
curves $k$ and $l$, depending on the energy difference $|\epsilon_k-\epsilon_l|$ and 
the velocity of the relative motion (see Sec.~\ref{nonab}). 
Here we assume that the relative motion is confined to the 
first $q$ eigenstates of $H_{\text{int}}$, i.e., there may be non-adiabatic transitions within 
the first $q$ eigenstates, but transitions to other states $\ket{\psi_l(\mf{R})}$ ($l>q$) can be neglected. 
In this case, projection of Eq.~(\ref{s1}) onto the  relevant levels  
$\mf{\mc{\tilde{\alpha}}} = (\alpha_1,\ldots,\alpha_q)$, $q < N$, results in the 
following Schr\"odinger equation for $\mf{\tilde{\alpha}}$~~\cite{dum:96,ruseckas:05,dalibard:11},
\begin{align}
\imath\hbar \partial_t \mf{\tilde{\alpha}} = \left[\frac{1}{2\mu}(\mf{p}\mathds{1} -\mf{\tilde{A}})^2 + \tilde{V} + \Phi \right] \mf{\tilde{\alpha}}. 
\label{s2}
\end{align}
Here $\tilde{V}$ and $\mf{\tilde{A}}$ are $q\times q$ matrices whose matrix elements for for $k,l\le q$ are 
given by Eqs.~(\ref{vdef}) and~(\ref{adef}), respectively. 
The scalar potential results from the remaining states outside the considered subspace and is defined as 
\begin{align}
\Phi_{kl} = \frac{1}{2\mu} \sum\limits_{p=q+1}^{N} \mf{A}_{kp}\cdot\mf{A}_{pl} .
\end{align}
The fields $\mf{\tilde{A}}$ and $\Phi$ transform under a basis change 
\begin{align}
\mf{\mc{\tilde{\alpha}}} & \rightarrow U(\mf{R}) \mf{\mc{\tilde{\alpha}}} 
\end{align}
according to 
\begin{align}
\mf{\tilde{A}} & \rightarrow U(\mf{R}) \mf{\tilde{A}} U^{\dagger}(\mf{R}) - \imath \hbar [\nabla U(\mf{R})]U^{\dagger}(\mf{R}), \\
\Phi & \rightarrow U(\mf{R}) \Phi U^{\dagger}(\mf{R}). 
\end{align}
It follows that $\mf{A}$ and   $\Phi$ can be regarded as gauge fields. 
The vector potential gives rise to an effective magnetic field $\mf{B}$ whose 
Cartesian components $B^{(i)}$ ($i\in \{1,2,3\}$) are defined as
\begin{align}
B^{(i)}& = \frac{1}{2} \ve_{ikl} F^{(kl)},  \label{bfeld}\\
F^{(kl)}& = \Omega^{(kl)}  -\frac{\imath}{\hbar}\left[\tilde{A}^{(k)},\tilde{A}^{(l)} \right],  \label{Ftensor}
\end{align}
where we employed Einstein's sum convention,  $\ve_{ikl}$ is the Levi-Civita tensor and 
\begin{align}
\Omega^{(kl)} & = \partial_k \tilde{A}^{(l)} -\partial_l \tilde{A}^{(k)} \label{bcurv} 
\end{align}
is the Berry curvature. Note that the last term in Eq.~(\ref{Ftensor}) 
is different from zero if the $q\times q$ matrices $\tilde{A}^{(k)}$ 
do not commute. In this case, the gauge fields are called non-Abelian. 
%
\begin{figure}[t!]
\includegraphics[width=6cm]{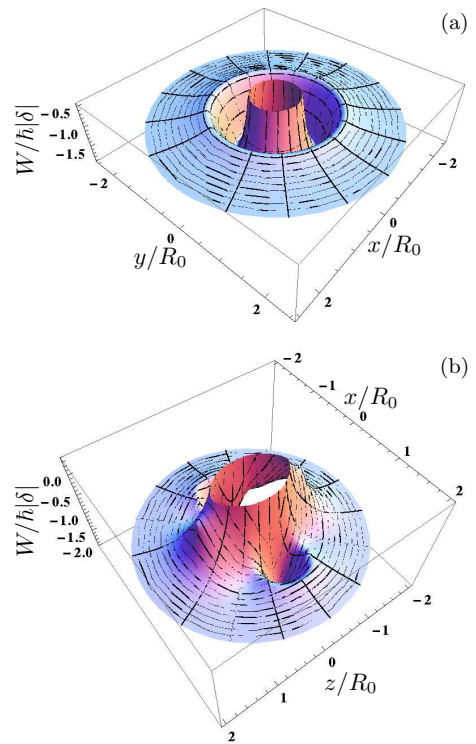}
\caption{\label{fig2}
(Color online) (a) Potential curve of the well state in the $x-y$ plane.  The potential is
azimuthally symmetric around the $z$ axis, giving rise to a donut-shaped potential well. 
 (b) Potential curve of the  well state in  the $x-z$ plane. 
Two pronounced minima occur on the $\pm x$ axes. 
In (a) and (b) we set $\Delta=-3|\delta|$. 
}
\end{figure}
%

%
In summary, the Hamiltonian in Eq.~(\ref{s2}) is equivalent to the minimal coupling Hamiltonian of a charged particle 
in an electromagnetic field, characterised by the vector potential $\mf{\tilde{A}}$ and scalar potential $\Phi$.  
We point out that   the charge of the fictitious particle with mass $\mu$
does not appear explicitly in Eq.~(\ref{s2}). All fields $\mf{\tilde{A}}$, $\mf{B}$ and 
$\Phi$ represent a  charge multiplied with the corresponding field. In our framework it is impossible 
to determine the value of the charge by auxiliary measurements, and hence its value is arbitrary. 
\section{Abelian gauge field   \label{abelian}} 
Here we consider the adiabatic motion  in a single eigenstate $\ket{\psi_1}$ of $H_{\text{int}}$ 
with energy $\epsilon_1$, see Eq.~(\ref{eigen}). Since this scenario corresponds to $q=1$ in Eq.~(\ref{s2}), 
the Cartesian components of the vector potential  are numbers and thus represent an 
Abelian gauge field. 
It was shown in~\cite{kiffner:12} that the two dipole-dipole interacting Rydberg atoms exhibit a 
donut shaped potential well in the $x-y$ plane for  $\delta=\Delta < 0$.  
We find that this potential well persists in the case of an asymmetric Stark splitting $\delta\not=\Delta$  
and denote the corresponding eigenstate and potential curve by $\ket{\psi_{\text{w}}}$ and 
$\epsilon_{\text{w}}$, respectively. 
In the following we set $\ket{\psi_1}=\ket{\psi_{\text{w}}}$, $\epsilon_1=\epsilon_{\text{w}}$ 
and investigate the vector potential, the corresponding magnetic field  and the spectrum of the 
Hamiltonian in Eq.~(\ref{s2}). 

\begin{figure}[t!]
\includegraphics[width=6cm]{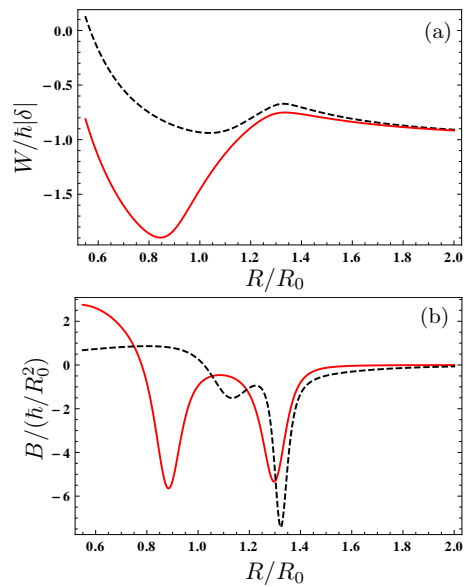}
\caption{\label{fig3}
(Color online) (a) Potential curve $\epsilon_{\text{w}}$ of the well state in the $x-y$ plane as a 
function of scaled internuclear spacing $R/R_0$. 
(b) The  $z$ component of the artificial magnetic field for $\mf{R}$ in the $x-y$ plane. In this 
spatial configuration, the $x$ and $y$ components of $\mf{B}$ are zero. 
The parameters in (a)-(b) 
are  $\Delta=-3|\delta|$ (solid red line) and $\Delta=-1.3|\delta|$ (black dashed line). 
}
\end{figure}

We begin with a characterisation of the potential  well  for $\delta \neq \Delta$. 
Due to the  azimuthal symmetry of the system,  a donut shaped potential well arises in the $x - y$ plane [see Fig.~\ref{fig2}(a)]. 
The spatial extend of the trapping potential in the $x - z$ plane  
is shown in Fig.~\ref{fig2}(b).  
The angular width of the potential around the $x$ axis is about $\pm 20^{\circ}$, which is similar to 
the case $\delta=\Delta$ studied in~\cite{kiffner:12}.  
The depth of the potential increases with decreasing ratio $\delta/\Delta$ of the detunings as shown in Fig.~\ref{fig3}(a).  
If the separation vector $\mf{R}$ lies in the $x-y$ plane, the well state $\ket{\psi_{\text{w}}}$ 
is contained within the $M=\pm 1$ manifold and its corresponding potential curve $\epsilon_{\text{w}}$ 
converges to the  $-\hbar |\delta|$ asymptote for $R\rightarrow\infty$. 
The potential minimum occurs roughly at 
$R_0$ [see Eq.~(\ref{r0})], which denotes the distance where the magnitude of the dipole-dipole interaction equals the 
Stark splitting $\hbar|\delta|$.

Next we investigate the vector potential  arising from the adiabatic motion in the well state $\ket{\psi_{\text{w}}}$, 
\begin{align}
\mf{A}_{\text{w}} =\imath\hbar \bra{\psi_{\text{w}}}\nabla \ket{\psi_{\text{w}}}. 
\end{align}
We find (see Appendix~\ref{fields}) that the only non-zero component is given by 
\begin{align}
A_{\text{w}}^{(\phi)}(\rho,z) = \frac{1}{\rho}\bra{\psi_{\text{w}}(\mf{R})} J_z \ket{\psi_{\text{w}}(\mf{R})}, \label{insightS}
\end{align}
where $J_z = J_z^{(1)} + J_z^{(2)}$ and $J_z^{(\mu)}$ is the $z$ component of the total angular momentum operator 
of the internal states of atom $\mu$.
Note that $A^{(\phi)}_{\text{w}}(\rho,z)$ does not depend on $\phi$ because of the azimuthal symmetry of 
the system. It follows that $\mf{A}_{\text{w}}$ obeys the Coulomb gauge ($\text{div} \mf{A}_{\text{w}}=0$) as 
a result of the phase convention described in Appendix~\ref{fields}.  
The magnetic field $\mf{B}=\nabla\times \mf{A}_{\text{w}}$ in Eq.~(\ref{bfeld}) is given by
\begin{align}
\mf{B} =  \frac{1}{\rho}\left[-\partial_z(\rho A^{(\phi)}_{\text{w}})\mf{e}_{\rho} + \partial_{\rho}(\rho A^{(\phi)}_{\text{w}}) \mf{e}_z\right] . 
\end{align}
Note that $\mf{B}$ can be also calculated independently of $A_{\text{w}}^{(\phi)}$ as shown in Appendix~\ref{curvature}.  
If the atoms are aligned in the $x-y$ plane, we find that only the component $B^{(3)}$ is different from zero. 
Figure~\ref{fig3}(b) shows $B^{(3)}$ for different ratios $\delta/\Delta$.

We emphasise that the magnetic field is only different from zero if we break the  
symmetry in the system, i.e., for  $\delta\not=\Delta$. In order to understand this result, we 
write the well state as 
\begin{align}
\ket{\psi_{\text{w}}} = \sum\limits_{M=-2}^{2} a_M \ket{p_M},
\end{align}
where $\ket{p_M}$ is the normalised projection of $\ket{\psi_{\text{w}}}$ onto the subspace with total angular momentum $M$. 
We thus have 
\begin{align}
\bra{\psi_{\text{w}}} J_z \ket{\psi_{\text{w}}}  = \hbar \sum\limits_{M=-2}^{2} M | a_M |^2 , 
\end{align}
and hence Eq.~(\ref{insightS}) implies that $A^{(\phi)}(\rho,z)$ will be zero if the level scheme is symmetric: In this case,  
$\ket{\psi_{\text{w}}}$ resides equally in states with positive and negative values of $M$, i.e., $|a_M | = |a_{-M}|$. 
Conversely, a broken  symmetry 
will give rise to a non-trivial vector potential since the population of $\ket{\psi_{\text{w}}}$ will be distributed unevenly 
over the $\pm M$ subspaces. 
%
%
%

%
The quantum dynamics of the relative atomic motion is governed by Eq.~(\ref{s2}). 
The corresponding Hamiltonian is given by 
\begin{align}
H_{\text{w}} = \frac{1}{2\mu}(\mf{p} -\mf{A}_{\text{w}})^2 + V_{\text{w}},
\label{zeeman1}
\end{align}
where  we  omitted the scalar potential $\Phi$. This is justified since it leads to a negligible 
modification of the potential $V_{\text{w}}$.  
Equation~(\ref{zeeman1}) represents the Hamiltonian of a spinless, charged particle with  mass $\mu$ 
in a magnetic field $\mf{B}$ and in the potential $V_{\text{w}}$. 
Next we investigate the bound states of $H_{\text{w}}$ and 
focus on the 2-dimensional setting where the motion is confined to the $x-y$ plane. 
This is justified because the dipole-dipole interaction confines the relative atomic motion 
to this plane, see Fig.~\ref{fig3}(b). 
We find that Eq.~(\ref{zeeman1}) can be written as 
\begin{align}
H_{\text{w}} = \frac{1}{2\mu}\mf{p}^2  + V_{\text{w}} + H_{\text{Z}} + H_{\text{D}},
\label{zeeman2}
\end{align}
where 
\begin{align}
H_{\text{Z}} & = - \frac{1}{\mu}\frac{A^{(\phi)}_{\text{w}}}{\rho} L_z , 
\label{zr} \\
H_{\text{D}} & =  \frac{\big[A^{(\phi)}_{\text{w}}\big]^2}{2\mu} . 
\label{dia}
\end{align}
%
\begin{figure}[t!]
\includegraphics[width=6cm]{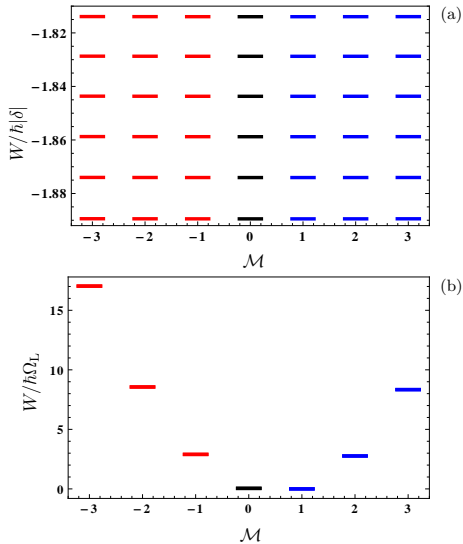}
\caption{\label{fig4}
(Color online) (a) Lowest vibrational states in the potential shown in Fig.~\ref{fig3}(a) for  $\Delta=-3|\delta|$ and 
$\Omega_{\text{L}}/|\delta|=2.8\times10^{-6}$. The energy difference of the vibrational 
states is $\omega_{\text{vib}}\approx 0.015 |\delta|$. 
(b) Lowest vibrational state for different quantum numbers 
$\mc{M}$ and on an energy scale defined by $\hbar\Omega_{\text{L}}$. 
}
\end{figure}
%
The operator $L_z$ in Eq.~(\ref{zr}) is the $z$ component of the orbital angular momentum associated with 
the relative atomic motion. It follows that $H_{\text{Z}}$ can be regarded as a  Zeeman Hamiltonian 
(see, e.g., complement DVII in~\cite{tannoudji:qm}).  
Note that the pre-factor $A^{(\phi)}_{\text{w}}/\rho$ in Eq.~(\ref{zr}) can be identified with 
$B_z$ if $A^{(\phi)}_{\text{w}}$ did not depend on $\rho$. 
The remaining contribution $H_{\text{D}}$ in Eq.~(\ref{dia}) is  the diamagnetic term of the Hamiltonian $H_{\text{w}}$ 
(see, e.g., complement DVII in~\cite{tannoudji:qm}). 

Next we show that the Zeeman Hamiltonian lifts the degeneracy between rotational states differing 
only by the sign of their azimuthal angular momentum. 
The splitting of Zeeman levels is typically given by the  Larmor angular frequency
$\omega_{\text{L}}= q B/(2m)$, where $q$ is the charge and $m$ is the mass of the particle.  
The corresponding quantity in our case is 
\begin{align}
\omega_{\text{L}}  = \frac{\hbar}{2\mu R_0^2}\tilde{B} = \Omega_{\text{L}} \tilde{B} , 
\end{align}
where $\tilde{B}=B/ (\hbar / R_0^2)$ is the dimensionless magnetic field shown in Fig.~\ref{fig3}(b) and 
\begin{align}
\Omega_{\text{L}}  = \frac{\hbar}{2\mu R_0^2}. 
\end{align}
The parameter $\hbar \Omega$ is the rotational constant of a rigid rotor comprised of two equal particles of mass 
$2\mu$ and separated by $R_0$. 
For ${}^{39}$K atoms with principal quantum number $n=30$ and $|\delta|= 2\pi \times 11.4\,\text{MHz}$, 
we obtain $R_0=2.85 \mu\text{m}$ and $\Omega_{\text{L}} \approx 2\pi \times 31.9 \, \text{Hz}$. 
Larger values can be obtained for lighter atoms and smaller values of $R_0$, which corresponds to 
lower values of the principal quantum number $n$. For example, 
for ${}^{23}$Na atoms with principal quantum number $n=15$ and $|\delta|= 2\pi \times 39.0\,\text{MHz}$, 
we obtain $R_0=0.75 \mu\text{m}$ and $\Omega_{\text{L}} \approx 2\pi \times 780 \, \text{Hz}$~\cite{comment}.

Next we find the spectrum of $H_{\text{w}}$ by numerical methods. Since $L_z$ commutes with 
all other parts of the Hamiltonian, we can find simultaneous eigenstates of $H_{\text{w}}$ and $L_z$. 
The eigenfunctions of $L_z$ are $e^{\imath \mc{M} \phi}$ with integer $\mc{M}$, and the corresponding eigenvalues are $\hbar \mc{M}$. 
Note that the quantum number $\mc{M}$ corresponds to the relative motion of the two Rydberg atoms, and must not 
be confused with the  azimuthal quantum number $m_{j}$ associated with the internal Zeeman states. 
Furthermore, we assume that $\alpha_{\text{w}}$ is strongly localised in the $z$ direction. 
The Ansatz 
\begin{align}
\alpha_{\text{w}}(\rho,z,\phi) \propto e^{\imath \mc{M}\phi} \mc{R}(\rho) \mc{Z}(z)
\end{align}
yields an effective equation for $\mc{R}(\rho)$, and we find the spectrum of the corresponding 
Hamiltonian via a discretisation of the spatial variable $\rho$.  
This procedure yields the vibrational energy states in the potential well in Fig.~\ref{fig3}(a). 
The result for the lowest vibrational states in the potential of Fig.~\ref{fig2} for  $\Delta=-3|\delta|$ 
is shown in Fig.~\ref{fig4}(a). 
We find that the energy difference of the vibrational 
states is $\omega_{\text{vib}}\approx 0.015 |\delta|$, which yields 
$\omega_{\text{vib}}\approx2\pi \times 171 \, \text{kHz}$ for $|\delta|= 2\pi \times 11.4\,\text{MHz}$. 
Each vibrational state is quasi-degenerate with respect to the quantum number $\mc{M}$ 
on the energy scale of $\hbar |\delta|$. 
However, Fig.~\ref{fig4}(b) shows the lowest vibrational state for different quantum numbers 
$\mc{M}$ and on an energy scale defined by $\hbar\Omega_{\text{L}}$. Here we clearly see the anticipated 
Zeeman splitting of the magnetic sublevels. The  splitting between the $\mc{M}=2$ and $\mc{M}=1$ state 
is approximately given by $2.76\times \Omega_\text{L}$. Note that this relation is approximately independent 
of the ratio $\Omega_{\text{L}}/|\delta|$. 
Since the dipole-dipole interaction is not a central potential, the total orbital angular momentum of the 
relative motion is not conserved. The kinetic energy term in 
$H_{\text{w}}$ contains a term that is quadratic in $\mc{M}$, 
and thus  the Zeeman spectrum looks different from 
that of atomic physics. 

We emphasise  that the ground state of the system is 
the $\mc{M}=1$ state [see Fig.~\ref{fig5}(b)], although the difference to the $\mc{M}=0$ state is small. 
In addition, all  $\mc{M}>0$ states lie below their negative counterparts. 
In order to provide a physical explanation for this result, we derive from Eq.~(\ref{zeeman1}) 
a dynamical equation for the Heisenberg operator $ \mf{R}$,
\begin{align}
\mu \partial_t^2 \mf{R} = & -\nabla  V_{\text{w}} \notag \\
& + \frac{1}{2\mu} \left\{\left[(\mf{p}-\mf{\tilde{A}})\times \mf{B}\right] 
- \left[\mf{B} \times  (\mf{p} - \mf{\tilde{A}})\right] \right\}. 
\label{lf}
\end{align}
The second line in Eq.~(\ref{lf}) describes the Lorentz force given by the vector 
product of the magnetic field and the  velocity  $(\mf{p}-\mf{\tilde{A}})/\mu$ 
of the relative motion. 
Since  the magnetic field points 
downwards  near the minimum of the potential well in Fig.~\ref{fig3}(a) [see Fig.~\ref{fig3}(b)], 
the Lorentz force favours  motion in anti-clockwise direction. This 
explains why the $\mc{M}>0$ states have lower energy as compared to their negative counterparts. 

The level structure of the bound states inside the potential well influence the quantum dynamics of the system. 
For example, it gives rise to a geometric phase which could be measured, in principle, via an Aharonov-Bohm type interference experiment~\cite{aharonov:59}. In addition, the deflection of the relative motion 
via the Lorentz force  offers a different route towards the experimental measurement 
of the effective magnetic field~\cite{prep}. 
\section{Non-Abelian gauge fields \label{nonab}}
\begin{figure}[t!]
\includegraphics[width=8.5cm]{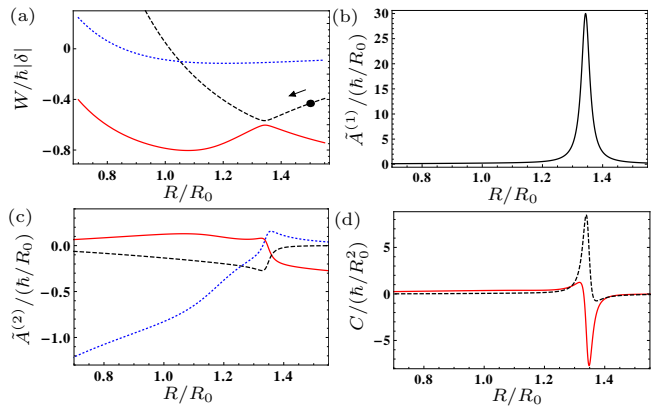}
\caption{\label{fig5}
(Color online) (a) Three potential wells in the $x-y$ plane corresponding to three different eigenstates of $H_{\text{int}}$. 
The solid red, black dashed, and blue dotted lines correspond to $\ket{\psi_1}$, 
$\ket{\psi_2}$, and $\ket{\psi_3}$, respectively. 
The black dot indicates the initial position of the system for the dynamics discussed in 
Fig.~\ref{fig6}, and the arrow indicates the direction of motion. 
(b) Imaginary part of  $\tilde{A}_{12}^{(1)}=[\tilde{A}_{21}^{(1)}]^*$ for $\phi=0$. 
(c) Real parts  of $\tilde{A}_{11}^{(2)}$ (solid red line), $\tilde{A}_{22}^{(2)}$ (dashed black line) and 
$\tilde{A}_{12}^{(2)}=[\tilde{A}_{21}^{(2)}]^*$ (dotted blue line) for $\phi=0$. 
(d) Matrix elements  of the commutator $C$ in Eq.~(\ref{com}). The red solid line shows 
$C_{11}=-C_{22}$, and the black dashed line represents $C_{12}=C_{21}$. 
In (a)-(d), we set $\Delta=-1.13|\delta|$. All components of $\mf{\tilde{A}}$ that are not shown in (b) and (c) are zero. 
}
\end{figure}

The vector potential $\mf{\tilde{A}}$ represents a non-Abelian gauge field if not all 
of its components $\tilde{A}^{(k)}$ commute with each other. This situation can only arise   if 
the relative motion induces non-adiabatic transitions between  $q\ge2$ eigenstates of $H_{\text{int}}$. 
In this case, the components of  $\mf{\tilde{A}}$ are $q\times q$ matrices that do not 
necessarily commute. 
Here we focus on the case $q=2$ and assume that $\ket{\psi_1}$ is the well state studied in 
Sec.~\ref{abelian}.  We find that the corresponding energy curve  $\epsilon_1$  exhibits 
an avoided crossing with  a second well state $\ket{\psi_2}$. 
This is shown in Fig.~\ref{fig5}(a), where $\epsilon_1$ ($\epsilon_2$) is represented by 
the solid red line (black dashed line). 
We point out that the two energy curves cross for $\delta=\Delta$, and 
hence the two lines represent a conical intersection~\cite{yarkony:96,domcke:ci} 
with respect to the parameters $\Delta/|\delta|$ and $R$. 
In the following we evaluate all components of the vector potential $\mf{\tilde{A}}$, 
and show that it can induce transitions between the states $\ket{\psi_1}$ and $\ket{\psi_2}$. 
Furthermore, we demonstrate that $\mf{\tilde{A}}$ is a non-Abelian gauge field. 
Note that the description of non-adiabatic transitions via gauge fields is 
a  well-established technique in molecules~\cite{yarkony:96,domcke:ci}. In addition, 
it has been applied to dipole-dipole interacting Rydberg atoms in~\cite{wuster:10,mobius:11,wuster:11}.

The  Hamiltonian for  the quantum dynamics of the relative motion and the two internal  states 
can be obtained from Eq.~(\ref{s2}) for $q=2$. We find 
\begin{align}
H_{2} = \frac{1}{2\mu}\left[\mf{p}^2\mathds{1}_2 -2\mf{\tilde{A}}\cdot\mf{p}-(\mf{p}\cdot\mf{\tilde{A}}) +\mf{\tilde{A}}^2\right] + \tilde{V},
\label{h2}
\end{align}
where  we  omitted the scalar potential $\Phi$ because its impact on the presented results is negligible.  
All off-diagonal terms of $H_2$  can give rise to 
a coupling between the relative atomic motion and the internal electronic states. 
In order to investigate the  coupling terms in Eq.~(\ref{h2}) in detail, 
we evaluate all components of the vector potential $\mf{\tilde{A}}$ as outlined in Appendix~\ref{fields}. 
The result is shown in Figs.~\ref{fig5}(b) and~(c), where all non-zero components of 
$\tilde{A}^{(1)}$ and $\tilde{A}^{(2)}$ are displayed, respectively. 
Note that we evaluate  $\mf{\tilde{A}}$  for $\phi=0$ such that 
$\tilde{A}^{(1)}$ ($\tilde{A}^{(2)}$) corresponds to the radial (azimuthal) component of $\mf{\tilde{A}}$.  
We find that the dominant coupling  term in Eq.~(\ref{h2}) is proportional to $\mf{\tilde{A}}_{12}\cdot\mf{p}$. 
Since the largest component of $\mf{\tilde{A}}_{12}$ is $\tilde{A}_{12}^{(1)}$ 
[see Figs.~\ref{fig5}(b) and~(c)], it follows that radial motion gives rise to the strongest coupling 
between states $\ket{\psi_1}$ and $\ket{\psi_2}$.  An estimate  shows 
that the coupling near the avoided crossing can become near-resonant for realistic Rydberg atom parameters. 
%
%
%

\begin{figure}[t!]
\includegraphics[width=6cm]{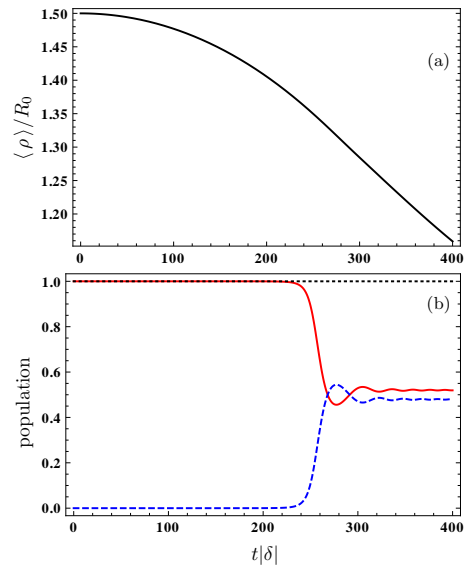}
\caption{\label{fig6}
(Color online) Semi-classical simulation of the system dynamics in the $x-y$ plane. The initial conditions are 
$\mean{\mf{\rho}}= 1.5 R_0$ and 
and $\mean{\mf{p}}=\mf{0}$. The system is  prepared in the internal state $\ket{\psi_2}$ at $t=0$.   
The parameters are $\Delta=-1.13|\delta|$ and $\Omega_{\text{L}}/|\delta|=2.8\times 10^{-6}$, 
corresponding to $|\delta|= 2\pi \times 11.4\,\text{MHz}$,  ${}^{39}$K atoms with principal quantum number $n=30$ and 
$R_0=2.85 \mu\text{m}$. 
We neglect effects due to the finite lifetime of the molecule ($t|\delta|\approx1300$)~\cite{comment2} and thus restrict 
the analysis to times that are significantly smaller. 
(a) Position $\mean{\rho}$ as a function of time. 
(b) Population of the internal states. The red solid line corresponds to $\ket{\psi_2}$, and the blue dashed line shows the population in 
$\ket{\psi_1}$. The black dotted line shows the sum of the population in  $\ket{\psi_1}$ and  $\ket{\psi_2}$. 
}
\end{figure}

%
A quantitative analysis of the coupling between internal states and the relative atomic motion 
can be obtained from a semi-classical simulation of the system dynamics. 
To this end we treat the relative motion of the atoms classically, i.e., we derive a set of 
coupled equations for the mean values $\mean{\mf{R}}$ 
and $\mean{\mf{p}}$ from the full Hamiltonian $H_{\text{int}}$ in Eq.~(\ref{Ht}). 
We assume that the dynamics is confined to the $x-y$ plane. 
This is justified because the anisotropic dipole-dipole interaction provides a trapping potential in the $z$ direction. 
Furthermore, we assume that the system is initially at rest and prepared in state $\ket{\psi_2}$ at $\rho=1.5 R_0$, 
see Fig.~\ref{fig5}(a). 
This could be realised if the ground state atoms are trapped in optical tweezers~\cite{gaetan:09,urban:09} and  
subsequently excited to the Rydberg and molecule levels. 
The result of the  semi-classical simulation of the system dynamics is shown Fig.~\ref{fig6}. 
At the initial position $\rho=1.5 R_0$ the system experiences the attractive part of the potential curve $\epsilon_2$. 
It thus  starts to move towards  the avoided crossing at $\rho\approx 1.33 R_0$, see 
Figs.~\ref{fig5}(a) and~\ref{fig6}(a). 
Near the avoided crossing transitions between the two internal states occur, and 
eventually an almost equal superposition of the states $\ket{\psi_1}$ and $\ket{\psi_2}$ is created 
[see Fig.~\ref{fig6}(b)]. 
Note that this semi-classical approach  does not account  for all physical phenomena.  
In particular, in a full quantum mechanical treatment the relative motion is described by a wavepacket 
that splits as it moves across the avoided crossing. The two wavepackets corresponding to the internal states 
$\ket{\psi_1}$ and $\ket{\psi_2}$ experience different potentials and will thus separate in space. 
It follows that the avoided crossing acts like a beamsplitter for the wavepacket of the relative motion. 
The full quantum mechanical analysis of this problem 
is beyond the scope of this work and will be presented elsewhere~\cite{prep}. 
The experimental observation of the splitting of the motional wavepacket requires 
measurements of the  density-density correlations of the two Rydberg atoms. 
Such measurements have been performed by ionization of the Rydberg atoms~\cite{schwarzkopf:11} 
and by de-excitation to the ground state followed by advanced imaging techniques~\cite{schauss:12}. 
We thus believe that the synthetic spin-orbit coupling is detectable with current or 
next-generation imaging techniques. 

The preceding results demonstrate that the vector potential $\mf{\tilde{A}}$ results in 
a coupling between the relative atomic motion and internal electronic states. It follows that 
our system realises a synthetic spin-orbit coupling. 
In addition, we find  that all components of the commutator 
\begin{align}
C & = \frac{\imath}{\hbar}  \left[\tilde{A}^{(1)},\tilde{A}^{(2)}\right]
\label{com}
\end{align}
are different from zero, see Fig.~\ref{fig5}(d). It follows that the gauge fields  $A^{(1)}$ and $A^{(2)}$ are non-Abelian. 
We point out that the diagonal elements of $C$ can be calculated directly without the knowledge of the 
vector potential $\mf{\tilde{A}}$. This is shown in Appendix~\ref{curvature}. 
The non-Abelian character of the gauge fields has a direct impact on the 
magnetic field via Eqs.~(\ref{bfeld}) and~(\ref{Ftensor}).  
Since the commutator $C$ is of the same order of magnitude as the artificial magnetic field 
experienced in state $\ket{\psi_1}$ alone [see Fig.~\ref{fig3}(b)], the non-Abelian 
effects will be of the same size as the impact of the magnetic field on 
the quantum dynamics in the internal state $\ket{\psi_1}$. 
In particular,  the Lorentz force acting on the relative atomic motion 
will contain a distinct signature of the non-Abelian 
character of the gauge fields~\cite{zygelman:90}. 
A more detailed investigation of non-Abelian signatures~\cite{jacob:07} in the quantum dynamics 
of our system is work in progress and will be presented elsewhere.

Finally, we point out that the full dynamics of this system is much richer than the $U(2)$ gauge theory discussed so far. 
The blue dotted line  in Fig.~\ref{fig5} shows the energy curve $\epsilon_3$ of a third state $\ket{\psi_3}$, which crosses $\ket{\psi_2}$ at $R\approx R_0$.  
This crossing turns into an avoided crossing outside the $x-y$ plane, and hence it can be 
regarded as a conical intersection~\cite{yarkony:96,domcke:ci}. 
For small values of $z$, the  coupling between $\ket{\psi_2}$ and 
$\ket{\psi_3}$ will be significant, and thus all three states have to be taken into account. 
%
%

\section{Summary and Discussion \label{summary}}
In summary, we have shown that the dipole-dipole interaction between Rydberg atoms 
can induce sizeable Abelian and non-Abelian artificial gauge fields affecting the 
relative motion of the two atoms. 
Our system is several orders of magnitude larger than conventional molecules, and 
the exquisite control over individual Rydberg atoms allows one to engineer non-trivial 
gauge fields. 
In the case of an Abelian gauge field and adiabatic motion in a single molecular state, 
the corresponding magnetic field results in a Zeeman shift of the rotational states. 
More specifically, the Zeeman shift lifts the degeneracy of rotational states whose 
azimuthal angular momenta have opposite sign. 
This result reflects the broken  symmetry in our system due to the 
different Stark shifts of the $m_j>0$ and $m_j<0$ magnetic sublevels of the individual 
Rydberg atoms. In particular, we find that the ground state of the potential well is given 
by the first excited rotational state. 

The relative atomic motion can induce transitions between internal two-atom states. This coupling 
is mediated by off-diagonal terms of the artificial gauge fields. We find that our system exhibits 
this synthetic spin-orbit coupling near an avoided crossing of two well states. In addition, we show 
that the corresponding gauge fields are non-Abelian. 

We expect that similar effects  arise in other dipole-dipole interacting 
systems like  polar molecules and magnetic atoms. Although the dipole-dipole interaction 
is much weaker in those systems, they are not limited by the finite lifetime of 
Rydberg atoms.

%
%

%
\appendix 
\section{Hamiltonian $H_{\text{int}}$ of the internal states \label{internal}}
The dynamics of the internal states is governed by the Hamiltonian $H_{\text{int}} = H_{\text{S}} + V_{\text{dd}}$. 
The Hamiltonian of the electronic levels $H_{\text{S}}$ is 
\begin{align}
H_{\text{S}} =   \hbar\sum\limits_{\mu=1}^{2} [& (\omega_0+\delta)\ket{p-1/2}_{\mu}\bra{p-1/2}_{\mu}  \notag  \\
 &  + (\omega_0+\Delta)\ket{p+1/2}_{\mu}\bra{p+1/2}_{\mu} \notag  \\
& + \omega_0(\ket{p+3/2}_{\mu}\bra{p+3/2}_{\mu} \notag \\
&+ \ket{p-3/2}_{\mu}\bra{p-3/2}_{\mu})]. 
\end{align}
Here $\omega_0$ is the resonance frequency  of the $\ket{s\pm1/2} \leftrightarrow \ket{p\pm3/2}$ transition. 

The symmetric level scheme with $\Delta = \delta$ can be realized if a 
static electric field is applied in $z$ direction~\cite{kiffner:12}. 
The asymmetry $\Delta\neq\delta$ could be induced, e.g., by the application of an additional 
$\sigma^+$ polarized AC field inducing AC Stark shifts~\cite{zimmerman:79,bohlouli:07} of the 
$\ket{s-1/2}\leftrightarrow\ket{p+1/2}$ and $\ket{s+1/2}\leftrightarrow\ket{p+3/2}$ transitions. 
Since the Clebsch-Gordan coefficients on these transitions are different, these Stark shifts change the 
energy difference between the $\ket{p+1/2}$ and $\ket{p+3/2}$ states and thus result in $\Delta\neq\delta$. 
In addition, the AC Stark shifts will lift the degeneracy of the $\ket{p\pm3/2}$ and $\ket{s\pm1/2}$ 
states. However, as shown in Sec.~\ref{abelian} any level scheme where states with 
positive and negative azimuthal quantum numbers have different energies results in an effective magnetic field. 
We have verified explicitly that non-degenerate $\ket{p\pm3/2}$ states result in qualitatively similar results 
as compared to the level scheme in Fig.~\ref{fig1}(a).

The dipole-dipole interaction  is given by~\cite{tannoudji:api}
\begin{align}
V_{\text{dd}} = \frac{1}{4\pi \epsilon_0 R^3}[\mf{\hat{d}}^{(1)}\cdot\mf{\hat{d}}^{(2)}
-3(\mf{\hat{d}}^{(1)}\cdot\vec{\mf{R}})(\mf{\hat{d}}^{(2)}\cdot\vec{\mf{R}})], 
\label{vdd}
\end{align}
where $\mf{\hat{d}}^{(i)}$ is the electric dipole-moment operator of atom $i$ and $\vec{\mf{R}}=\mf{R}/R$ 
is the unit vector along the molecular axis. 
For the evaluation of Eq.~(\ref{vdd}) we compute the matrix elements of the dipole operator 
via  the Wigner-Eckart theorem~\cite{edmonds:amq}  according to 
\begin{align}
 \langle p m |\mf{\hat{d}}^{(i)}|s m' \rangle = 
\underbrace{\frac{(np_{3/2}\|\mf{\hat{d}}\|ns_{1/2})}{[2\cdot3/2 +1]^{1/2}}}_{=\mathcal{D}} 
\ \sum_{q=-1}^1 C^{3/2 m}_{1/2 m' 1 q} \vec{\mf{\epsilon}}_q, 
\label{mele}
\end{align}
where  $C^{3/2 m}_{1/2 m' 1 q}$ are Clebsch-Gordan coefficients 
and $\vec{\mf{\epsilon}}_q$ are orthonormal unit vectors 
arising from the decomposition of the dipole operator into its spherical components~\cite{edmonds:amq}. 
The reduced matrix element  $(np_{3/2}\|\mf{\hat{d}}\|ns_{1/2})$ in Eq.~(\ref{mele}) 
can be written in terms of a radial matrix element between the  $np_{3/2}$ and $ns_{1/2}$ states~\cite{edmonds:amq,park:11,walker:08,kiffner:12}. 
Since the sum in Eq.~(\ref{mele}) is a term of order unity, the magnitude of $V_{\text{dd}}$  is given by 
\begin{align}
\hbar \Omega=|\mathcal{D}|^2/(4\pi \epsilon_0 R^3). 
\end{align}
Equating $\Omega$ to $|\delta|$ yields the characteristic length $R_0$, given by
\begin{align}
R_0=(|\mathcal{D}|^2/(4\pi \epsilon_0\hbar |\delta|)^{1/3}.
\label{r0}
\end{align}
We scale all energies with the Stark shift $\hbar |\delta|$, and thus $V_{\text{dd}}$ only depends on the ratio $R/R_0$. 
In the subspace of $nsnp$ states, the frequency $\omega_0$ appears as an offset to all states and can thus be eliminated. 
Scaling  $H_{\text{S}}$  with $\hbar |\delta|$ leaves the ratio $\Delta/\delta$ as the only remaining free parameter.

\section{Evaluation of the vector potential $\mf{A}$ \label{fields}}
The off-diagonal elements of the vector potential are given by~\cite{xiao:10}
\begin{align}
\mf{A}_{kl} & = \frac{\imath \hbar}{\epsilon_l-\epsilon_k}\bra{\psi_k(\mf{R})}[\nabla H_{\text{int}} 
(\mf{R})]\ket{\psi_l(\mf{R})} . 
\label{vecA}
\end{align}
Note that Eq.~(\ref{vecA}) can be evaluated without any phase convention for the eigenstates $\ket{\psi_l(\mf{R})}$, i.e., via the 
numerical diagonalisation of $H_{\text{int}}$. 
On the other hand, Eq.~(\ref{vecA}) will only yield continuous expressions for $\mf{A}_{kl}(\mf{R})$ as a function of $\mf{R}$ if a 
suitable phase  convention for the eigenstates $\ket{\psi_l(\mf{R})}$ is imposed.
In the following, we employ a particular phase convention that allows us 
to evaluate the diagonal elements of $\mf{A}$ as well. 

We first explain our phase convention for states in a plane with constant $\phi$. Without loss of generality we consider the 
plane with $\phi=0$ and pick one reference point $\mf{R}_0$. In order to  fix the phase of an eigenstate $\ket{\psi_n(\mf{P})}$ 
at $\mf{P}$ we impose the condition of parallel transport, i.e., 
\begin{align}
\bra{\psi_n(\mf{R})}\nabla\ket{\psi_n(\mf{R})}\cdot \mf{e}_{\mf{R}_0\mf{P}} = 0, 
\label{para1}
\end{align}
where the unit vector $\mf{e}_{\mf{R}_0\mf{P}}$ points from $\mf{R}_0$ to $\mf{P}$ and $\mf{R}$ is an arbitrary point on the straight line connecting 
$\mf{R}_0$ and $\mf{P}$. In the absence of degeneracies, this means that we obtain $\ket{\psi_n(\mf{P})}$ by adiabatic evolution 
from $\ket{\psi_n(\mf{R}_0)}$. 
In general, the  condition~(\ref{para1}) does not guarantee that the phase of $\ket{\psi_n(\mf{R})}$ is a smooth 
function of position. If one considers adiabatic evolution on a closed path in the $\phi=0$ plane, the state will acquire 
a Berry phase that is equal to the magnetic flux enclosed by the circular path. However, the nature of the dipole-dipole interaction 
cannot create a magnetic flux in $\mf{e}_{\phi}$ 
direction, and thus Eq.~(\ref{para1}) leads to a well-defined phase as a function of position. 
In order to put this line of thought on formal grounds, 
we prove that the magnetic field component $B_{n}^{(\phi)}$ is indeed zero for every state $\ket{\psi_n}$. 
With Eq.~(\ref{Fq1}) we find 
\begin{align}
[(\nabla\times \mf{A}_n)\cdot \mf{e}_{\phi}]_{\phi=0} & = (\partial_3 A_n^{(1)} - \partial_1 A_n^{(3)})_{\phi=0}  \label{lastl} \\
& = \frac{\imath}{\hbar}\sum\limits_{p\not=n} \left(A_{np}^{(3)}A_{pn}^{(1)} -A_{np}^{(1)}A_{pn}^{(3)}\right)_{\phi=0} \notag \\
& = \frac{\imath}{\hbar}\sum\limits_{p\not=n}2i\, \text{Im}\left(A_{np}^{(3)}A_{pn}^{(1)} \right)_{\phi=0} . \notag
\end{align}
Equation~(\ref{vecA}) allows us to write 
\begin{align}
A_{np}^{(3)} A_{pn}^{(1)} = \frac{\hbar^2}{(\epsilon_p-\epsilon_n)^2}
\bra{\psi_n}[\partial_3 H_{\text{int}}]\ket{\psi_p} \bra{\psi_p}[\partial_1 H_{\text{int}}]\ket{\psi_n} .
\label{phi0}
\end{align}
We find that $\partial_3 H_{\text{int}}$, $\partial_1 H_{\text{int}}$ and $H_{\text{int}}$ are real and  symmetric 
matrices for $\phi=0$. 
In particular, since  $H_{\text{int}}$ is real and symmetric,  all  components of the eigenstates $\ket{\psi_n}$ in the Zeeman basis are 
real [up to an overall complex factor, which drops out in Eq.~(\ref{phi0})]. 
It follows that the expression in Eq.~(\ref{phi0}) is real, and thus we have  
\begin{align}
[(\nabla\times \mf{A}_n)\cdot \mf{e}_{\phi}]_{\phi=0} = \partial_z A_n^{(\rho)} - \partial_r A_n^{(z)} =0.  \label{ephi0}
\end{align}
This implies that there exists a function $S_n(\rho,z)$ such that $\partial_{\rho}S_n = A_n^{(\rho)}$ and $\partial_{z}S_n = A_n^{(z)}$. 
Condition~(\ref{para1}) means that the line integral of $\mf{A}_n$ from $\mf{R}_0$ to $\mf{P}$ is zero, 
and hence $S_n$ is constant throughout the $\phi=0$ plane. It follows that  
\begin{align}
\partial_{\rho}S_n & =A_n^{(\rho)}=\bra{\psi_n(\mf{R})}\nabla\ket{\psi_n(\mf{R})}\cdot \mf{e}_{\rho}=0, \label{para2a} \\
\partial_{z}S_n & = A_n^{(z)} = \bra{\psi_n(\mf{R})}\nabla\ket{\psi_n(\mf{R})}\cdot \mf{e}_z = 0 
\label{para2b}
\end{align}
at every point in the $\phi=0$ plane. 

Since the system exhibits azimuthal symmetry, all states at $\mf{R}$ outside the plane with $\phi=0$ can be 
generated from those in the plane via a unitary rotation operator~\cite{kiffner:07b}, 
\begin{align}
\ket{\psi_n({\rho},z,\phi)} = e^{-\imath/\hbar J_z \phi} \ket{\psi_n({\rho},z,\phi=0)},
\label{rotateS}
\end{align}
where $J_z = J_z^{(1)} + J_z^{(2)}$ and $J_z^{(\mu)}$ is the $z$ component of the total angular momentum operator 
of the internal states of atom $\mu$. It follows from Eq.~(\ref{rotateS}) that relations~(\ref{para2a}) and~(\ref{para2b}) hold for any $\mf{R}$, and 
the only non-zero component of $\mf{A}_n$ is given by 
\begin{align}
A_n^{(\phi)}(\rho,z) & =  \imath\hbar \bra{\psi_n(\mf{R})}\nabla\ket{\psi_n(\mf{R})}\cdot \mf{e}_{\phi} \\
&  = \imath\hbar  \frac{1}{{\rho}}\bra{\psi_n(\mf{R})}\partial_{\phi} \ket{\psi_n(\mf{R})}  . 
\end{align}
According to Eq.~(\ref{rotateS}), $A_n^{(\phi)}(\rho,z)$ is proportional to the mean value of the operator $J_z$, 
\begin{align}
A_n^{(\phi)}(\rho,z) = \frac{1}{{\rho}}\bra{\psi_n(\mf{R})} J_z \ket{\psi_n(\mf{R})}. 
\end{align}
This result allows us to evaluate the diagonal elements of $\mf{A}_n$ by numerical means. 
%
%
\section{Evaluation of the curvature \label{curvature}}
The Berry curvature in Eq.~(\ref{bcurv}) can be written as 
\begin{align}
\Omega_{nm}^{(kl)}  & = \imath \hbar \left(\braket{\partial_k\psi_n}{\partial_l\psi_m}-\braket{\partial_l\psi_n}{\partial_k\psi_m}\right) \\
& = \frac{\imath}{\hbar} \sum\limits_{p=1}^{N} \left( A_{np}^{(k)} A_{pm}^{(l)} - A_{np}^{(l)} A_{pm}^{(k)} \right) \label{OmS}\\
& = \frac{\imath}{\hbar} \left[A^{(k)},A^{(l)} \right]_{nm} ,  \label{OmS2}
\end{align}
where we employed Eq.~(\ref{vecA}).
We point out that  Eqs.~(\ref{Ftensor}) and~(\ref{OmS2}) imply that the magnetic field is equal to zero if all $N$ states comprise the system of interest, 
i.e., if $q=N$. If the system of interest consists of the first $q$ states, we have 
\begin{align}
F_{nm}^{(kl)} = \Omega_{nm}^{(kl)} -\frac{\imath}{\hbar}\left[\tilde{A}^{(k)},\tilde{A}^{(l)} \right]_{nm},
\end{align}
where $n,m \le q$ and $\tilde{A}^{(k)}$ is a $q\times q$ matrix. The commutator on the right hand side of the latter equation can be written as 
\begin{align}
\frac{\imath}{\hbar}\left[\tilde{A}^{(k)},\tilde{A}^{(l)} \right]_{nm} 
= \frac{\imath}{\hbar}\sum\limits_{p=1}^{q}\left(A_{np}^{(k)}A_{pm}^{(l)} -A_{np}^{(l)}A_{pm}^{(k)}\right),
\end{align}
and thus  Eq.~(\ref{OmS}) yields 
\begin{align}
F_{nm}^{(kl)} = \frac{\imath}{\hbar}\sum\limits_{p>q}^{N}\left(A_{np}^{(k)}A_{pm}^{(l)} -A_{np}^{(l)}A_{pm}^{(k)}\right).
\label{ften}
\end{align}
Since Eq.~(\ref{ften}) contains only non-diagonal elements of $\mf{A}$, it can be evaluated via the result in Eq.~(\ref{vecA}). 
Moreover, we emphasise that all diagonal elements of $F_{nn}^{(kl)} = F_{n}^{(kl)}$ are completely independent of the phase of the eigenstates. 
It follows that the diagonal elements of $\mf{B}$ can be calculated via the numerical diagonalisation of $H_{\text{int}}$. 
In particular, for $q=1$ we have 
\begin{align}
F_{1}^{(kl)} = \Omega_{1}^{(kl)}=\frac{\imath}{\hbar} \sum\limits_{p>1} \left( A_{1p}^{(k)} A_{p1}^{(l)} - A_{1p}^{(l)} A_{p1}^{(k)} \right). 
\end{align}
Note that  the latter equation holds for any of the internal states if we  re-arrange their order. 
More generally, we thus have 
\begin{align}
F_{n}^{(kl)} = \Omega_{n}^{(kl)}=\frac{\imath}{\hbar} \sum\limits_{p\not=n} \left( A_{np}^{(k)} A_{pn}^{(l)} - A_{np}^{(l)} A_{pn}^{(k)} \right). \label{Fq1}
\end{align}
Finally, we demonstrate that the above results can be employed for the (numerical) 
evaluation of the diagonal elements of the commutator $[\tilde{A}^{(k)},\tilde{A}^{(l)}]$. 
We have 
\begin{align}
F_{n}^{(kl)} & = \Omega_{n}^{(kl)} \quad \text{for} \quad q=1 . \label{q1}\\
F_{n}^{(kl)} & = \Omega_{n}^{(kl)} -\frac{\imath}{\hbar}\left[\tilde{A}^{(k)},\tilde{A}^{(l)} \right]_{nn}\quad \text{for} \quad q > 1 . \label{qb}
\end{align}
Subtraction of Eq.~(\ref{qb}) from~(\ref{q1}) yields
\begin{align}
\frac{\imath}{\hbar} \left[\tilde{A}^{(k)},\tilde{A}^{(l)} \right]_{nn} =  F_{n}^{(kl)}(q=1) - F_{n}^{(kl)}(q>1)  
\end{align}
Since $F_{n}^{(kl)}(q=1)$ and  $F_{n}^{(kl)}(q>1)$ can be evaluated numerically without any phase convention for the eigenstates, 
the same is true for the diagonal elements of the commutator $[\tilde{A}^{(k)},\tilde{A}^{(l)}]$. 


\begin{thebibliography}{10}
\bibitem{comment}%
  \BibitemOpen
  \bibinfo {note} {Note that $|\delta|$ must be much smaller than the $np$
  fine-structure splitting $\omega_{\text{FS}}$~\cite{kiffner:12}. Throughout
  this paper we choose $ 50 |\delta|\le \omega_{\text{FS}} $.}%
  \bibAnnoteFile{Stop}{comment}%
\bibitem{comment2}%
  \BibitemOpen
  \bibinfo {note} {The lifetime of a Rydberg dimer with $n=40$ was estimated
  in~\cite{kiffner:12} to be $42\,\mu s$ for alkali atoms. Since the lifetime
  scales with $n^3$, we find $t|\delta|\approx1300$ for $|\delta|= 2\pi \times
  11.4\,\text{MHz}$ and $n=30$.}%
  \bibAnnoteFile{Stop}{comment2}%
\end{thebibliography}
\end{document}